\documentclass{llncs}
\usepackage{epsfig}
\usepackage{amssymb}
\usepackage{amsmath}
\usepackage{times}
\usepackage{graphicx}
\usepackage{latexsym}
\usepackage{algorithm}
\usepackage{algorithmic}
\usepackage[colorlinks]{hyperref}

 \newcommand{\NI}{\noindent}

\begin{document}

\title{Similarity Assessment through blocking and  affordance assignment in Textual CBR}
\titlerunning{Affordance Assessment in TCBR}          
\author{R.Rajendra Prasath\thanks{This work was carried out during the tenure of an ERCIM ``Alain
Bensoussan'' Fellowship Programme.} \ and \ Pinar \"{O}zt\"{u}rk}

\authorrunning{Rajendra and Pinar}                  
\tocauthor{Rajendra Prasath, Pinar \"{O}zt\"{u}rk}  

\institute{Department of Computer and Information Science (IDI)\\
Norwegian University of Science and Technology (NTNU), \\
Sem S{\ae}lands Vei 7-9, NO - 7491, Trondheim, Norway, \\
\email{\{rajendra,pinar\}@idi.ntnu.no},\\
\texttt{\url{http://www.idi.ntnu.no/\~rajendra};
\\\url{http://www.idi.ntnu.no/people/pinar};} }

\date{}                      
\maketitle                   
\thispagestyle{empty}        

\begin{abstract}
It has been conceived that children learn new objects through
their affordances, that is, the actions that can be taken on them.
We suggest that web pages also have affordances defined in terms
of the users' information need they meet. An assumption of the
proposed approach is that different parts of a text  may not be
equally important / relevant to a given query. Judgment on the
relevance of a web document requires, therefore, a thorough look
into its parts, rather than treating it as a monolithic content.
We propose a method to extract and assign affordances to texts and
then use these affordances to retrieve the corresponding web
pages. The overall approach presented in the paper relies on
case-based representations that  bridge the queries to the
affordances of web documents. We tested our method on the tourism
domain and the results are promising.
\end{abstract}

\section{Introduction}
 \label{s:Intro}
 \NI

World Wide Web (WWW) is a massively distributed and decentralized
medium for information and services, and also one of the most
egalitarian discoveries of mankind in modern times.  However, the
use of the web technology to its maximum possible extent  requires
development of flexible and  effective searching approaches. To
this end, we propose an approach in which the web can be searched
through case representations   that capture plausible connections
between users' queries and affordances of web documents.

This work presents an approach to document retrieval in the
tourism domain, yet the underlying research objective is to
develop a method for explication and capture of the {\it
affordance} of documents that are available on the WWW.
Gibson\cite{Gibson77} introduced the term {\it affordance} to
refer to the opportunities for action provided by a certain {\it
object}. We suggest that  web documents should similarly have
affordances that refer to use-purposes of documents. The meaning
of a content to the query of a user lies in its affordance. The
question transforms then into how the affordance of documents are
conveyed in textual format and can be extracted and represented in
a reusable way.

A central idea of the presented method is that a document may
contain information that matches different queries, each of which
corresponds to an affordance. For example, a web document about
{\it New Delhi} may provide information about the non-vegetarian
restaurants, the bazaars, as well as the transportation within the
city while the main focus may be on shopping. This would mean that
most of the content revolves around bazaars, shopping malls, and
the shopping norms (i.e., whether/where to bargain and when/where
not).  The main strategy of the approach is to divide the Web
pages into text segments, determine affordance of each and use
these to determine the information needs the documents afford.
This may be considered as a special type of {\it tagging}
technique, in the classical natural language processing
terminology.

\begin{figure}[htbp]
 \vspace*{-.6cm}
 \centerline{\psfig{figure=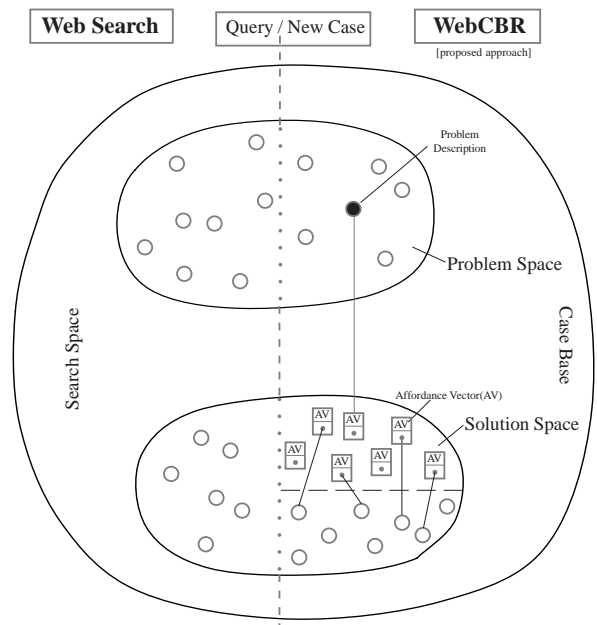,width=3.4in,height=3.4in}}
 \vspace*{-.6cm}
 \caption{Blocking and affordance assignment approach in WebCBR}
 \label{figBT01}
\end{figure}

The use of Case Based Reasoning (CBR) in web context has attracted
researchers for a while. For example, Limthan $et$
al.~\cite{Limthanmaphon03} applied CBR in order to compose a
complex web service from heterogeneous web services residing in
different parts of the web, and \cite{Diaz2006} in order to search
and select web services. Ha~\cite{Ha08} investigated how the web
usage data can be used to discover navigation patterns which, in
turn, can be used to predict the user behavior. In this work, a
web document has a corresponding case representation capturing
information that  bridges queries to documents through
affordances.  Query experiences are  used  to adjust the
affordances of documents. This  provides  search-useful  and
re-usable information for future web searches. The key rationale
of the proposed approach is based on two assumptions:  {\it (i)} a
web document embraces a number of text blocks each of which can be
connected to one or several affordances,  {\it (ii)} the web can
be searched CBR-wise  and relevance of a document can be judged on
the basis of its affordance alignment with the current query case.
Figure.~\ref{figBT01} illustrates the proposed approach in
comparison with the web search.

The paper is organized as follows: Next section presents the idea
behind the web-affordance notion. Section~\ref{s:PropWork}
explains the proposed case-based approach and the case population.
Section \ref{s:expResults} presents the experimental results and
discussions. Finally, section~\ref{s:Concln} wraps up with
conclusions and future directions.

\section{The affordance-guided querying approach}
 \label{s:phil}

A main rationale behind the proposed approach is the hypothesis
that a document may serve more than one information need, that is,
it may have multiple affordances. This because different parts
(here we consider them as {\it blocks})  of a document may have
slightly different focuses. We defined a list of topics in advance
(manually at the moment) each of which fulfills an information
need of the user. In tourism domain,  a user may need information
related to accommodation at a particular place, easy and
economical transport options, places to roam around, time to
travel from one place to another place, food / types of
restaurants, historical / important places to visit, shopping at
famous places, so on and so forth. Currently we have a list of 18
topics for the tourism domain. Each web page  may afford to one or
several of such information needs.

Affordance of a text segment/block is represented as a vector of
which each element specifies the extent the text affords a certain
information need in the affordance list for the task domain. The
affordance of a whole document is determined on the basis of the
affordance vectors (AV) of its constituent blocks. The size of an
affordance vector, consequently, is $m$ in the example (ie.,
tourism) domain (here $m$ = 18).

The querying approach proposed in this paper relies on retrieving
a number of documents relevant to the query using information
retrieval (IR) techniques, and then employing an affordance-based
ranking  on this set. In the rest of the paper, we use the
following notation:

\vspace{-0.5cm}
 \begin{equation}
  \label{e:AV}
 AV_{\rm text}=  \{ A_1, A_2, ........, A_m\}
 \end{equation}

\noindent where {\it AV} is a vector while ${\it A_i}$ is a scalar
representing the affordance with respect to the $i^{\rm th}$
element in the list of {\it m} affordances.

\section{Case-Based, Affordance-Guided Web}
 \label{s:PropWork}
 \NI

The objective of this work is, given a query, to retrieve the web
documents that meet the user's information need in the best
possible way,  using a refined assessment of the document guided
by affordances. The underlying assumption is that the web is
informed about the affordance of each document, in a case base.
There is a case for each document of which the problem description
part consists of a term-set  that represents the document in a
concise way, and the identification of the document it represents.
The term-set is, in a sense, a dimension-reduced version of the
web page. The solution part informs about the affordance of the
document indirectly, through affordances of its constituent
blocks. Hence a case is represented as follows:

$C_i$= $\{ProbDescription, Solution\}$

\noindent where {\it ProbDescription} consist of a   term-set
representation of the web page while {\it Solution} has two
components: \vspace{.1cm}

$Solution$= $\{ AV, I \}$

\noindent  {\it AV} is  the affordance vector of size {\it m} (see
Equation  \ref{e:AV}) and represents the affordances of the web
page. AV, together with {\it I} ( which is the identification of
the corresponding document), constitutes the {\it solution} part
of a case. The next section describes how the case base is
populated.

\subsection{Population of the Case Base}
 \label{ss:casePopul}
 \NI

Initially, case base contains no cases and cases are constructed
incrementally in the off-line mode. To build the case base, we
apply {\it link-to-text} ratio which is defined as the ratio
between the size of the text tagged with hyperlinks and the text
without hyperlinks.

\noindent {\bf Problem description}: A web page is first segmented
into blocks and a  textual description of each block $b_{i}$ is
extracted (after removing the markups and stop words) using {\it
link-to-text} ratio.  If {\it link-to-text} ratio is more then,
all hyperlinked text will be skipped and {\it link-to-text} ratio
is scaled in the normalized interval [0-1]. Then from each
extracted block text, top $k$ discriminative terms are selected
(after stop word removal) and added to {\it ProbDescription} in
the problem description of the case. This process is described in
the `problem part' of Algorithm 1.

\noindent {\bf Solution}: The  extracted text, say {\it
text}$_{i}$, of the block $b_{i}$  is processed to identify its
affordance with the help of the resource terms (topics and \#
terms considered in each topic, are presented in the
table~\ref{t:affordanceTerms}). For each topic, the matching terms
are identified, the affordance with respect to this topic is
computed and the affordance vector of the block is updated (see
{\bf ComputeBlockAffordance} in the Algorithm 2) accordingly. The
AV of the document, in turn, is computed by  {\bf
ComputeDocAffordance}, as described in the `solution part' of the
Algorithm 1. A case is generated by coupling the problem
description and the solution parts and is added to the case base.

\vspace*{-.6cm}
\begin{algorithm}[htbp]
 \label{PopulateCB}
 \caption{Population of Case Base}

\vspace*{.1cm}
 {\bf Input:} \ \ List of topic: $L_{t}$ (from table~\ref{t:affordanceTerms}) ;

 \hspace*{1cm} A web document: $d$;

 \vspace*{.1cm}
 {\bf Procedure:}

 I: identification of  $d$;

 {\bf Problem Part:}
\begin{algorithmic}[1]
        \STATE Initialize probDesc to NULL;
        \FOR {each block text data}
            \STATE remove stop words and punctuation;
            \STATE filter out top $k$ words from the block text
            and add it to probDesc;
        \ENDFOR
\end{algorithmic}

{\bf Solution Part:}

\begin{algorithmic}[1]
        \STATE $AV_d = {\bf ComputeDocumentAffordance}(d)$
    \STATE Soln = $<AV_d, I >$

\end{algorithmic}

{\bf Case Base:}

\begin{algorithmic}[1]
    \STATE Add the new case having $<$probDesc, Soln$>$ to the case base;
\end{algorithmic}

\vspace*{.1cm}

{\bf Output:}
\smallskip The {\rm Case Base};
\end{algorithm}
\vspace*{-1.0cm}

\subsection{The Querying Process}
 \label{ss:retrieval}
 \NI

Two main processes underlying the querying method are described in
Algorithm 3. This part is similar to information retrieval for the
given query, but retrieval is performed on a specific amount of
extracted text from each block and the assigned affordances.
During the retrieval, the problem description part of the cases
are matched with the user's query and top k cases are retrieved.
These are then ranked using the solution of the case, where the AV
of the query and the AV of cases are compared. AV of the query is
computed on the basis of the terms (also called as `resource
terms') in each topic. After each such a retrieval episode, the AV
of  the top {\it k} number of cases are revised and modified in
such a way that its currently experienced relevance to the query
is properly reflected in the case representation.

 \vspace*{-0.4cm}
\begin{algorithm}[htbp]
 \label{AlgoGetDocAffordance} 
 \caption{Procedures to compute Document and  Block Affordance Vectors}

 \vspace*{.1cm}
 {\bf Procedure: ComputeDocAffordance($d$)}

 \vspace*{.1cm} {\bf Input:} \ List of topics - $L_{t}$; \ \ A web document - $d$;

 \vspace*{.1cm}
 {\bf Procedure:}

\begin{algorithmic}[1]
    \STATE Segment $d$ into blocks $b_{i}$
    Initialize $AV_{d}$:= NULL;
    \FOR {each block $b_{i}$ in $d$}
        \STATE Compute $AV_{b_{i}} := {\bf ComputeBlockAffordance}(b_{i})$
        \STATE Update $AV_{d} := AV_{d} + AV_{b_{i}}$
    \ENDFOR

    \vspace*{.1cm}

    \STATE return $AV_d$
\end{algorithmic}

{\bf Output:}
\smallskip The {\rm AV} of the given document $b_{i}$

 {\bf ------------------------------------------------------------------------------------------------------------------}
 \vspace*{.1cm}
 {\bf Procedure: ComputeBlockAffordance($b_{i}$)}

\vspace*{.1cm}
 {\bf Input:} \ List of topics - $L_{t}$ (as in table~\ref{t:affordanceTerms}); \ \ $AV_{b_{i}}$ - affordance vector;

 \vspace*{.1cm}
 {\bf Procedure:}

\begin{algorithmic}[1]
    \FOR {each {\it topic}$_{j} \in L_{t}$}
        \STATE Compute the number of matching terms in $b_{i}$ and terms in {\it topic}$_{j} \in L_{t}$
        \STATE Update this score for the corresponding affordance in
        $AV_{b_{i}}$.
    \ENDFOR
    \STATE return $AV_{b_{i}}$
\end{algorithmic}

\vspace*{.1cm}

{\bf Output:}
\smallskip The {\rm affordance vector} $AV_{b_{i}}$ of the block
$b_{i}$
\end{algorithm}
\vspace*{-.7cm}

\begin{algorithm}[htbp]
\label{AlgoSAusingBT} \caption{Similarity Assessment through
blocking and affordance assignment}

\vspace*{.1cm}
 {\bf Input:} \ \ A query  having $n$ terms: $q = \{ q_{t_{1}}, q_{t_{2}}, \cdots , q_{t_{n}}\}$

 \hspace*{1cm} Case base having $m$ cases: $\{ c_{1}, c_{2}, \cdots , c_{m}\}$

 \hspace*{1cm} $L_{t}$ - List of topics;

 \hspace*{1cm} ${AV}_{q}$ - the query affordance vector

 \hspace*{1cm} ${AV}_{c}$ - the affordance vector of  a case.

 \vspace*{.1cm}
 {\bf Procedure:}

\begin{algorithmic}[1]
    \STATE Retrieve top $k$ cases using
        $$sim(q, c_{j}) = \sum\limits_{\substack{t_{i} \in q; \\ t_{j} \in c_{j}; \\ {\rm matching \ terms} } } sim(t_{i}, t_{j})$$
        where $q$ is the query; $c_{j}$ is the case and $t_{j}$ are the
        matching terms in the problem part of $c_{j}$
    \FOR {each retrieved case $c_{k}$}
        \STATE compute the query affordance vector ${AV_{q}}$ with respect
        to  the topic  list $L_{t}$;
        \STATE get ${AV_{c}}$ from the solution of $c_{k}$;
        \STATE compute $sim({AV_q}, {AV_c})$ using cosine metric;
        (see equation \ref{eqn:simEst})
    \ENDFOR
    \STATE return the ranked list of top $k$ cases with respect to
    $sim({AV_q}, {AV_c})$
\end{algorithmic}

{\bf Output:}
\smallskip The ranked list of cases sorted by their similarity scores
\end{algorithm}

\section{Experimental Results}
 \label{s:expResults}
 \NI
 \vspace*{-1cm}

\subsection{Web Corpus}
 \label{ss:Dataset}
 \NI

The effectiveness of the proposed method is analyzed through
experimental results on a corpus containing the web pages mostly
related to the tourist places in India. The tourism web pages were
collected by applying the crawling process according to a set of
policies that filter the supplementary files. We have omitted the
web pages having only hyperlinks, images, advertisements and
graphical layouts (like the index page of the most of the sites).
Additionally we skipped the pages containing redirect options,
less significant textual description, only copyright information,
etc. The remaining pages are collected to form a raw web corpus.
Then preprocessing tasks were performed to generate the case base
having problem description and solution parts through content and
structure mining with focused information extraction.

\subsection{Preprocessing}
 \label{ss:preProcess}
 \NI

We have applied focused content filtering which performs the
structural mining on each collected web page. This structural
mining, based on \textsf{table} OR \textsf{paragraph} OR
\textsf{div} tags, decomposes the given web page into blocks. Then
for each block, we applied the {\it link-to-text} ratio to
distinguish content noise and content text description. We perform
duplicate sentences elimination both at the phrase level and at
the whole sentence level in order to avoid repeating sentences in
the solution parts of each block text. The extracted text content,
if they are represented in hex code, are converted into unicode.
So multilingual content using hex representation can also be
processed (except for the pages using certain proprietary fonts /
encodings). We have retained the headings and paragraph markers
with selected top $k$ terms for the problem description. But due
to link to text ratio, some of the headings might have been
removed. Specific patterns are hardly seen for eliminating the
unlinked noise from such pages.

Among the total number of 112,522 web documents [1315 seed URLs
were crawled to the depth 3], 14,033 web pages, containing both
tourism and non tourism pages but having sufficient textual data
(after filtering the web pages having spam contents like unwanted,
restricted contents, adult contents, etc), were selected for our
experiments. We have manually crafted the list of 18 affordances
related to tourism domain including the affordance {\it
miscellaneous}.

\begin{table}[htbp]
\centering \vspace*{-0.5cm}
\begin{tabular}{|l|c|l|c|}
 \hline 
 Affordances & \# Terms &  Affordances & \# Terms \\ \hline
 Accommodation  & 59 & Retreats       & 59 \\ \hline
 Attractions    & 59 & Shopping       & 59 \\ \hline
 Beaches        & 59 & Spirituality   & 59 \\ \hline
 Deserts        & 59 & Sports         & 66 \\ \hline
 HealthCare     & 60 & ThemeParks     & 59 \\ \hline
 Heritage       & 59 & TourPackages   & 59 \\ \hline
 HillStations   & 59 & Transport      & 61 \\ \hline
 Landscapes     & 59 & Wildlife       & 59 \\ \hline
 Nature         & 59 & Miscellaneous  & Rest \\ \hline
\end{tabular}
\caption{List of predefined affordances with number of terms in
each affordance}
 \label{t:affordanceTerms}
 \vspace*{-1.2cm}
\end{table}

\subsection{Queries / New Cases}
 \label{ss:query}
 \NI

We have considered 25 tourism queries in English language used in
Cross Lingual Information Access (CLIA)
Project\footnote{\url{http://www.clia.iitb.ac.in/clia-beta-ext/}}
- a large project on cross-lingual information access systems for
Indian languages, that is being funded by the Government of India,
and being executed by a consortium of several academic
institutions and industrial partners\cite{Mandar01}. Each query is
presented in three forms: \textsf{title} - the actual query,
\textsf{desc} - the expanded query and \textsf{narr} - the
narration of the query. At present, we have attempted with
\textsf{title}, \textsf{desc} parts. Here we considered each query
( \textsf{title} / \textsf{desc}) as a new case.

\subsection{Evaluation Methodology}
 \label{ss:EvalMethod}
 \NI

In the experiments, we compared the effectiveness of the retrieval
using Lucene\footnote{\url{http://lucene.apache.org/java/docs/}}
and the proposed approach. In Lucene, the similarity scoring
function\cite{LuceneMan} is derieved from its conceptual formula
as follows:
\begin{eqnarray}
sim(q, d) &:=& coord(q,d) \cdot queryNorm(q) \\ & & \cdot \sum_{t
\in q} ( tf(t \in d) \cdot idf(t)^2 \cdot t.getBoost() \cdot
norm(t,d) ) \nonumber
\end{eqnarray}
\NI where $tf(t \in d)$ is the term frequency - the number of time
$t$ occurs in $d$; $idf(t)$ is the inverse document frequency of
the term; $ coord(q,d)$ is the score factor based on how many of
the query terms are found in the specified document;
$queryNorm(q)$ is the normalizing factor used to make scores
between queries comparable (does not affect the document ranking);
$t.getBoost()$ is the field boost and $norm(t,d)$ encapsulates a
few (indexing time) boost and length factors. Here {\it norm}
values is encoded during index time and decoded during search
time. Thus encoding/decoding comes with the precision loss - that
means $decode(encode(x)) = x$ is not guaranteed. Lucene allows the
users to customize its scoring formula by changing the boost
factors for calculating the score of similarity between the query
$q$ and the document $d$. Lucene\cite{LuceneMan} sorts the
retrieved results based on either their {\it relevance} or {\it
index order} for the given query. Here we sort the retrieved
results based on their relevance to the given query.

In the proposed approach, we weight the AV of the case in the
similar way to \cite{BuckleySAS94,SinghalSMB96}:

Weighted affordance of a case($W_{c_i}$):
\begin{equation}
\frac{w_{c_i}}{\sqrt{\sum_{i = 1}^{m} w_{c_i}^2}}
\end{equation}
where $w_{c_i}$ is the weight of the affordance $i$ in the AV of
the case.

Weighted affordance of a Query(new case)($W_{q_i}$):
\begin{equation}
 \frac{w_{q_i}}{\sqrt{\sum_{i = 1}^{m} w_{q_i}^2}}
\end{equation}
where $w_{q_i}$ is the weight of the affordance in the AV of the
query.

Similarity between the query (new case) and the case is computed
by:
\begin{equation}
\label{eqn:simEst}
 sim({q_i}, {c_i}):=  sim(AV_{q_i}, AV_{c_i}):= \sum_{\rm matching \ features} W_{q_i} \times W_{c_i}
\end{equation}

During the estimation of case affordance vectors, the values of
the elements in the vector increase with the number of matching
terms between the solution part and the new case. In such
situations, we could apply affordance vector length normalization.
To length normalize the elements of the affordance vector
$AV_{c}$, for the case $c$ having $m$ affordances: $\{ A_{t_1},
A_{t_2}, \cdots, A_{t_m} \}$, to the unit vector, we do the
following: $AV_{c} : = AV_{c} / |AV_{c}|$ where denominator
denotes the Euclidean length of the vector of the affordance
${t_i}$ in $c$. In the mean time, we will take care of the effect
of normalization factor in decreasing the chances of retrieval of
the document.
 \vspace*{-.2cm}
\begin{figure}[htbp]
 \centerline{\psfig{figure=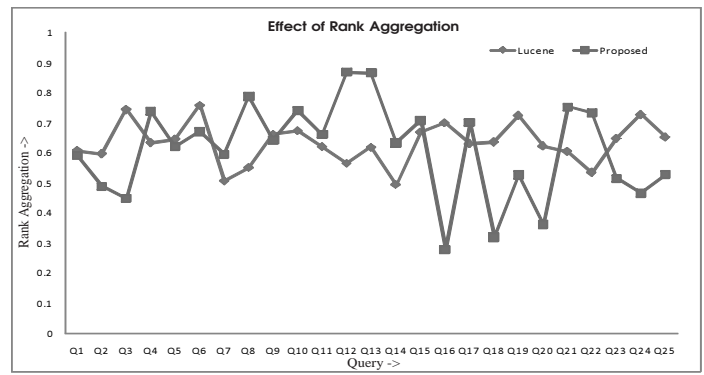,width=3.8in,height=2.0in}}
 \caption{Effect of rank aggregation of lucene retrieval vs the proposed approach}
 \label{figBT02}
\end{figure}
 \vspace*{-.4cm}

\begin{figure}[htbp]
 \centerline{\psfig{figure=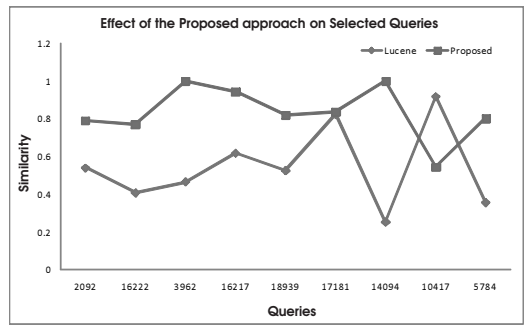,width=2.8in,height=1.8in}}
 \vspace*{-.5cm}
 \caption{Effect of the proposed approach on a few selected queries}
 \label{figBT03}
 \vspace*{-.5cm}
\end{figure}

We perform {\it rank aggregation}: Given a new case, retrieve top
candidate cases using the problem description. Then similarity
estimates of the affordance vector (solution part) of each of the
candidate cases with the affordance vector of the query are
computed. Finally the rank of all top $k$ cases are aggregated
with respect to their actual similarity scores and the results are
compared. The figure~\ref{figBT02} shows the effects of rank
aggregation for the 25 tourism queries  [\textsf{title} /
\textsf{desc} parts are considered here]. Lucene retrieval score
is influenced by index time boot factors and applies tf - idf
tradeoff with overall content of the textual description. In the
proposed approach, the point of focus is the affordance assignment
with respect to the block text based on its maximum affordance.
The queries Q12, Q13 and Q14 are proper names representing the
places and each extracted block text, that speaks about these
places specifically, contributes to the overall affordance.
Similarly Q21 and Q22 focus on the specific event / hotel in the
particular place. This gives combined affordance score with with
the proper names. Hence this leads to a better performance for
most of the queries (particularly for Q12, Q13, Q21 and Q21).

Next we considered a few sample queries whose similarities are
effectively computed through blocking and affordance
assignment(fig.~\ref{figBT03}). For example, in Q1 (Query:
\textsf{sunderbans national park}), the document with ID - 2092,
ranked 14 in Lucene retrieval and the proposed method has brought
it to rank 2. Here affordances related to {\it wildlife}, {\it
heritage} and {\it attractions} are captured where as the
affordances related to {\it nature} is hardly captured. This is
due to the fact that \textsf{park}  under the affordance
\textsf{nature}  contributed less to the overall affordance than
to \textsf{wildlife}. In another example, for the query
\textsf{Elephant Safari in Kaziranga}, the document, with ID:
10417, having the dominating term of ``safari'', has been brought
to the top in lucene where as its affordance value score very
less. At the same time, for the query \textsf{Goddess Meenakshi
Temple}, the proposed approach captured the document, with ID:
5784, whose blocks describe different topics related to Meenakshi
temple in Madurai, Tamil Nadu, India.

Even though the performance of the proposed system is promising,
the effect of the noise and the accuracy of filtering approach
along with the list of resource terms play a vital role in the
effective retrieval of cases. The effect of spam pages will reduce
the chances of retrieving the relevant document through boosting
their scores by projecting the related themes. Owing to paucity of
resources, we have limited our spam filtering to filter pages
containing adult content along with tourism related textual
content. This effectively reflects in the retrieved results with
the proposed approach. This is our preliminary attempt with
manually crafted term list for each identified (predefined)
affordance related to the tourism domain. Developing an automated
process for the affordance identification irrespective of the
domain may be attempted in the future.

\section{Conclusion}
 \label{s:Concln}
 \NI

We presented an approach for achieving an effective case retrieval
through the similarity assessment based on blocking and
identifying affordances of web documents. Affordance provides a
visual clue to the case identification. Traditional methods
dealing with textual content try to apply similarity metrics
collectively for the heterogeneous blocks of text together
presented in the same web content. This reduces the chances of the
user expected contents (solutions). The proposed approach solves
this issue by applying page segmentation through blocks,
identifying valid text from these blocks, scoring each of the
blocks with certain affordance scores and then applying similarity
metrics to achieve the effective case retrieval. The actual
performance of the proposed approach can be seen with the
similarity scores computed using query affordance vector and case
affordance vectors. The preliminary results show that the proposed
approach would be promising for identifying the specific block
text as the relevant solution.

\subsubsection*{Acknowledgement}
 \NI

{\small We gratefully acknowledge: Prof. Sudeshna Sarkar (IIT
Kharagpur) for initiating this work; the efforts of Mr. Parin and
Mr. Gyan (IIT, Kharagpur) in getting the list of seed urls for
crawling and Mr. David Kabath (NTNU) in preparing the list of
(tourism related) keywords. Also authors thank the anonymous
reviewers for their insights which improved the quality of the
paper}

\bibliographystyle{abbrv}

\end{document}